\documentclass[11pt,a4paper]{article}
\usepackage{jinstpub}

\makeatletter
\@ifpackageloaded{natbib}{}{\usepackage{cite}}

\@ifpackageloaded{hyperref}{}{\usepackage[colorlinks,citecolor=blue,urlcolor=blue]{hyperref}}

\@ifundefined{oldSI}{%
\usepackage[separate-uncertainty,retain-explicit-plus,per-mode=symbol,product-units=power]{siunitx}
}{%
\usepackage[separate-uncertainty,retain-explicit-plus,per-mode=symbol,product-units=power]{siunitx}[=v2]
\newcommand\qty{\SI} 
\newcommand\qtyproduct{\SI}
\newcommand\qtyrange{\SIrange}
}
\makeatother

\DeclareSIUnit\cps{cps}
\DeclareSIUnit\electron{\mbox{$e^-$}}
\DeclareSIUnit\barg{barg}
\DeclareSIUnit\oV{OV}
\usepackage{lineno}
\usepackage{isotope}

\usepackage[utf8]{inputenc}
\usepackage[T1]{fontenc} 
\usepackage{appendix}
\usepackage{todonotes}
\usepackage{amsmath}
\usepackage{wasysym}
\usepackage{nicefrac}
\usepackage{tabularx}

\DeclareMathSymbol{\mdot}{\mathord}{symbols}{"01}

\usepackage{multirow}
\usepackage{caption}
\usepackage{subcaption}
\notoc

\title{Very large SiPM arrays with aggregated output}

\newcommand{\AQLNGS}{INFN Laboratori Nazionali del Gran Sasso, Assergi (AQ) 67100, Italy}
\newcommand{\AQGSSI}{Gran Sasso Science Institute, L'Aquila 67100, Italy}

\newcommand{\BOINFN}{INFN Bologna, Bologna 40126, Italy}

\newcommand{\Princeton}{Physics Department, Princeton University, Princeton, NJ 08544, USA}

\newcommand{\RHUL}{Department of Physics, Royal Holloway University of London, Egham TW20 0EX, UK}

\newcommand{\TNFBK}{Fondazione Bruno Kessler, Povo 38123, Italy}
\newcommand{\TNTIFPA}{Trento Institute for Fundamental Physics and Applications, Povo 38123, Italy}

\newcommand{\UCLA}{Physics and Astronomy Department, University of California, Los Angeles, CA 90095, USA}

\author[a]{A.~Razeto,}
\author[a]{V.~Camillo,}
\author[a]{M.~Carlini,}
\author[a]{L.~Consiglio,}
\author[b]{A.~Flammini,}
\author[c,d]{C.~Galbiati,} 
\author[a]{C.~Ghiano,}
\author[e,f]{A.~Gola,}
\author[c]{S.~Horikawa,} 
\author[c]{P.~Kachru,} 
\author[a]{I.~Kochanek,} 
\author[a]{K.~Kondo,}
\author[g,a]{G.~Korga,}
\author[e,f]{A.~Mazzi,}
\author[c]{A.~Moharana,}
\author[e,f]{G.~Paternoster,}
\author[a]{D.~Sablone,}
\author[h]{H.~Wang}

\affiliation[a]{\AQLNGS}
\affiliation[b]{\BOINFN}
\affiliation[c]{\AQGSSI}
\affiliation[d]{\Princeton}
\affiliation[e]{\TNFBK}
\affiliation[f]{\TNTIFPA}
\affiliation[g]{\RHUL}
\affiliation[h]{\UCLA}

\emailAdd{sarlabb7@lngs.infn.it}

\abstract{In this work we will document the design and the performances of a SiPM-based photo-detector with a surface area of \SI{100}{\square\cm} conceived to operate as a replacement for PMTs. The signals from \num{94} SiPMs are summed up to produce an aggregated output that exhibits in liquid nitrogen a dark count rate (DCR) lower than \SI{100}{cps} over the entire surface, a signal to noise ratio better than 13, and a timing resolution better than \SI{5.5}{\nano\second}. The module feeds about \SI{360}{\milli\watt} at \SI{5}{\volt} with a dynamic range in excess of \num{500} photo-electrons on a \SI{100}{\ohm} differential line. The unit is compatible with operations at room temperature, with a DCR increased by about \num{6} orders of magnitude.}


\begin{document}
\maketitle

\section{Introduction}
Very large scale experiments are being studied or under construction to unlock the fundamental properties of our universe.
Experiments such as Dune, DarkSide, XenonNT, and Darwin share the requirement of detecting faint pulses of light with many photo-sensors capable of operating in cryogenic environment~\cite{dune, ds20k, xenonnt, darwin}.

In the past~\cite{tile}, we demonstrated that it is possible to tile \num{24} large SiPMs to build a photo-detector working in liquid nitrogen/argon with unprecedented performances. In this work we present a \qtyproduct{100x100}{\mm} photo-detector with all the auxiliary electronic components required for installation in a particle detector. 

\section{SiPMs}
In this work NUV-HD-Cryo SiPMs from FBK have been used; these devices represent an evolution from ~\cite{cryo-nuv} in terms of stability at cryogenic temperature. This allows for higher cell size and lower quenching resistance with respect to the NUV-HD-LF devices used in~\cite{tile, cryo-pre}. An overview of the SiPM electrical parameters is reported in Table~\ref{tab:sipm}. The performances of NUV-HD-Cryo SiPMs have been reported in~\cite{nuv-hd-cryo, 2pac, star}: in liquid nitrogen, the Dark Count Rate (DCR) is lower than \qty{1}{\cps\per\square\cm} at \qty{9}{\oV} with a negligible after-pulsing probability (\qty{<10}{\percent}), while the internal cross-talk can reach \qty{50}{\percent} at the highest over-voltages.

\begin{table}[t]
\centering
\begin{tabular}{llcc}
\hline
\textbf{Group} & \textbf{Parameter} at \qty{77}{\kelvin} & \textbf{NUV-HD-Cryo} & \textbf{NUV-HD-LF} \\ \hline
\multirow{7}{*}{SiPM} & SiPM Size & \multicolumn{2}{c}{\qtyproduct{7.9x11.7}{\mm}}\\
 & Cell Unit Size ($S_{c}$) & \qty{30}{\micro\meter} & \qty{25}{\micro\meter}\\ 
 & Cell Capacitance ($C_{d}$) & \qty{65}{\femto\farad} & \qty{40}{\femto\farad}\\  
 & Number of cells ($N_{c}$) & \qty{100}{k} & \qty{150}{k} \\  
 & Quenching Resistance ($R_{q}$) & \qty{3\pm0.3}{\mega\ohm} & \qty{10\pm1}{\mega\ohm}\\ 
 & Breakdown Voltage ($V_{bd}$) & \qty{27.5\pm0.3}{\volt} & \qty{28\pm0.4}{\volt} \\
 & Maximum Over-voltage ($OV_{MAX}$) & 9 V & 6 V \\
 & Primary Recharge Time ($R_{q} \times C_{d}$) & \qty{180}{\nano\second} & \qty{400}{\nano\second} \\ \hline
\multirow{4}{*}{\begin{tabular}[c]{@{}l@{}}4$\times$2s3p Tile\end{tabular}} & Aggregated Recharge Time ($\tau$) & \qty{350}{\nano\second} & $\approx$\qty{600}{\nano\second} \\ 
 & Current Peak ($I_{p}$ = ½ $C_{d}$ / $\tau$) & \qty{100}{\nano\ampere\per\oV} & \qty{30}{\nano\ampere\per\oV} \\ 
 & Input noise density at \qty{1}{\mega\hertz} & \qty{18}{\pico\ampere\per\sqrt{\hertz}} & \qty{15}{\pico\ampere\per\sqrt{\hertz}} \\ 
 & SNR with Matched Filter & \qty{4}{\per\oV} & \qty{4}{\per\oV} \\ \hline
\multirow{4}{*}{\begin{tabular}[c]{@{}l@{}}4s6p Tile\end{tabular}} & Aggregated Recharge Time ($\tau^{+}$) & \qty{350}{\nano\second} & \\ 
 & Current Peak ($I_{p}$ = ¼ $C_{d}$ / $\tau$) & \qty{50}{\nano\ampere\per\oV} &  \\ 
 & Input noise density at \qty{1}{\mega\hertz} & \qty{9}{\pico\ampere\per\sqrt{\hertz}} & \\
 & SNR with Matched Filter & \textgreater \qty{3.8}{\per\oV} &  \\ \hline
\end{tabular}
\caption{Electrical parameters of the SiPMs and of the assembled tiles in liquid nitrogen. The first section documents the electrical parameters of the single SiPMs: the NUV-HD-LF column refers to the devices studied in~\cite{cryo-nuv, cryo-pre, tile}, while the NUV-HD-Cryo are the subject of this work and have been documented in~\cite{nuv-hd-cryo, 2pac, star}. For the breakdown voltage and for the quenching resistance the error term describes the standard deviation of the population, rather than a measurement uncertainty. The second section compares the performances of the assembled tile with the 4$\times$2s3p ganging scheme described in~\cite{tile}, while the third section describes the performance of the new 4s6p ganging scheme described in this work.}
\label{tab:sipm}
\end{table}

\section{4s6p Tile}
The peak output current of the NUV-HD-Cryo SiPMs (defined as $OV / R_q$) is more than three times larger than what was generated in the same condition by the NUV-HD-LF devices, Table~\ref{tab:sipm}.
The higher signal permits the use of a ganging topology that is oriented towards the reduction of the dissipated power at the expense of a lower current at the input of the trans-impedance pre-amplifier. 

\subsection{Ganging scheme}
\begin{figure}[tb]
\centering
\begin{subfigure}{.63\textwidth}
\includegraphics[width=\textwidth, ]{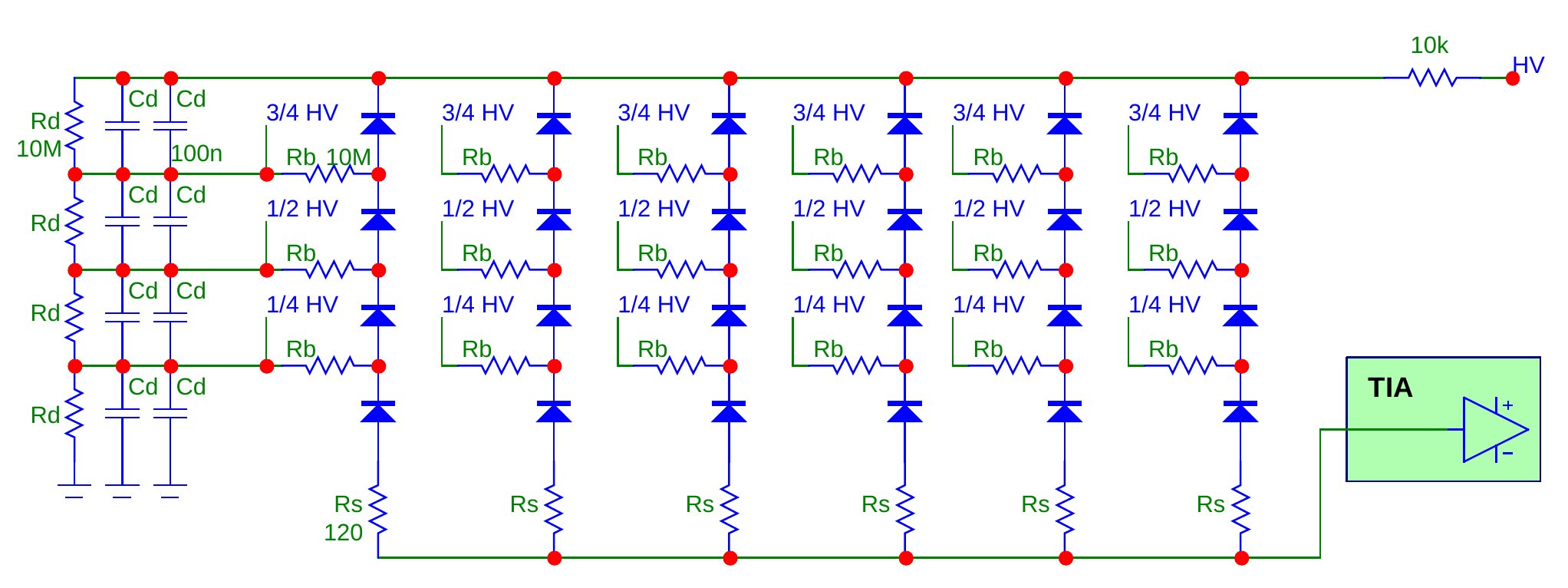}
\caption{4s6p Tile}
\end{subfigure}
\hfill
\begin{subfigure}{.35\textwidth}
\includegraphics[width=\textwidth, ]{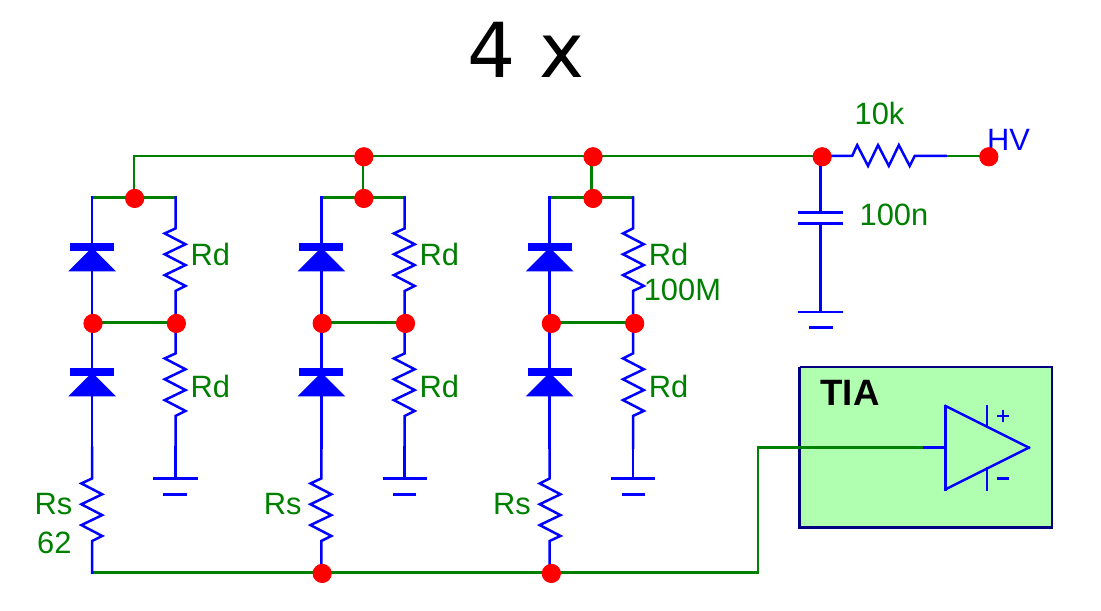}
\caption{2s3p Quadrant}
\end{subfigure}
\caption{Schematics of the 4s6p SiPM layout: six branches of four SiPMs are connected with a single TIA pre-amplifier for a total of 24 sensors. As a reference, one quadrant of the 4$\times$2s3p tile is shown: the readout of 24 SiPMs is achieved by summing four identical stages with an active adder (not shown).}
\label{fig:4s6p}
\end{figure}

Figure~\ref{fig:4s6p} depicts the \textit{4s6p} topology where six branches each with four SiPMs are fed to a single TIA.
In~\cite{tile} we used four quadrants each with an individual TIA to read 24 SiPMs: each quadrant was configured with a \textit{2s3p} layout (three parallel branches of two SiPMs in series) and an active summing node was required to aggregate the signal from the four TIAs.

The 4s6p configuration drastically simplifies the scenario: only one pre-amplifier is required without other circuit elements. However, the stronger ganging reduces the signal by a factor two with respect to the 2s3p scenario. This happens because the capacitive coupling of the SiPMs in series attenuates the photo-current by the series order (4 in this case). The next section will demonstrate the possibility to maintain nearly the same signal to noise ratio as that of the 4$\times$2s3p configuration. This is achieved by reducing the input noise of the TIA with a proper selection of the circuit elements.

\subsection{Divider}
A precision voltage divider is required to provide an even voltage bias distribution to all the devices. At cryogenic temperature, the DCR-induced current is in sub-picoampere, a region that is sensitive to surface leakages of the SiPMs or from the PCB.
The passive divider, shown in Figure~\ref{fig:4s6p}, is provided by four resistors.
The accuracy of the bias distribution affects the quality of the signal. Assuming the operation at \qty{7}{\oV} in liquid nitrogen, a disuniformity of \qty{1}{\percent} in the resistor leads to an over-voltage spread of \qty{5}{\percent}. This is reflected in the form of an equivalent variation of the gain between different SiPMs in the same tile. 
The value of $R_d$ defines the stability of the divider for current leakages of the SiPMs or of the PCBs: with $R_d$=\qty{10}{\mega\ohm} a SiPM surface leakage of \qty{50}{\nano\ampere} (over \qty{30}{\volt}) would affect the divider by less than \qty{1}{\percent}.

For the choice of $R_d$, Vishay resistors MCT06030C1005FP500 have been selected as they provide a \qty{\pm1}{\percent} tolerance with very low temperature drift.
Eight \qty{100}{\nano\farad} PEN capacitors (Panasonic ECW-U1104V33) are placed in parallel to the divider. The capacitors and the \qty{10}{\kilo\ohm} resistor form a filter for the noise coming from the bias line.
The branch resistors $R_b$ decouple the SiPMs from each other and from the divider capacitors, thus eliminating unwanted paths to the signal. For convenience, same Vishay part for the branch resistors and for the divider resistors were used.

\begin{figure}[tb]
\centering
\begin{subfigure}{0.41\textwidth}
\includegraphics[width=\textwidth, ]{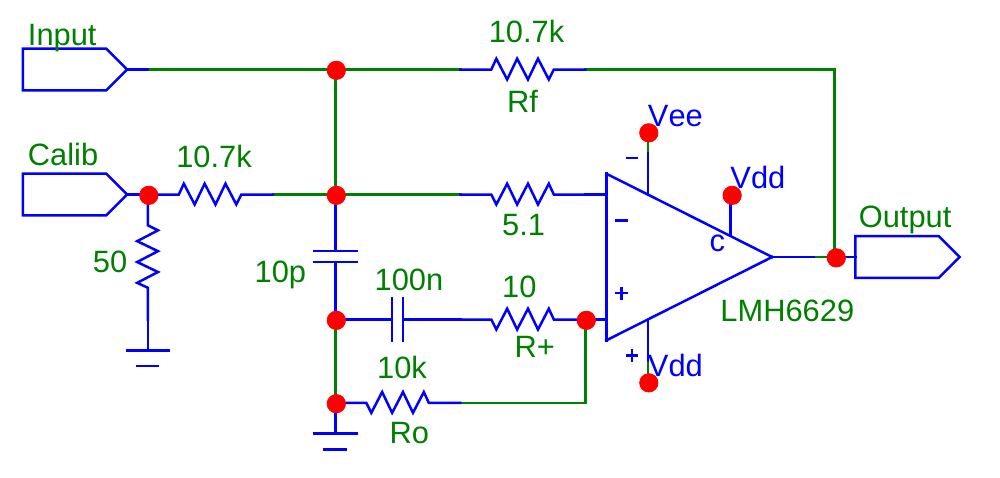}
\caption{Schematics}
\label{fig:tia}
\end{subfigure}
\hfill
\begin{subfigure}{0.58\textwidth}
\includegraphics[width=\textwidth, ]{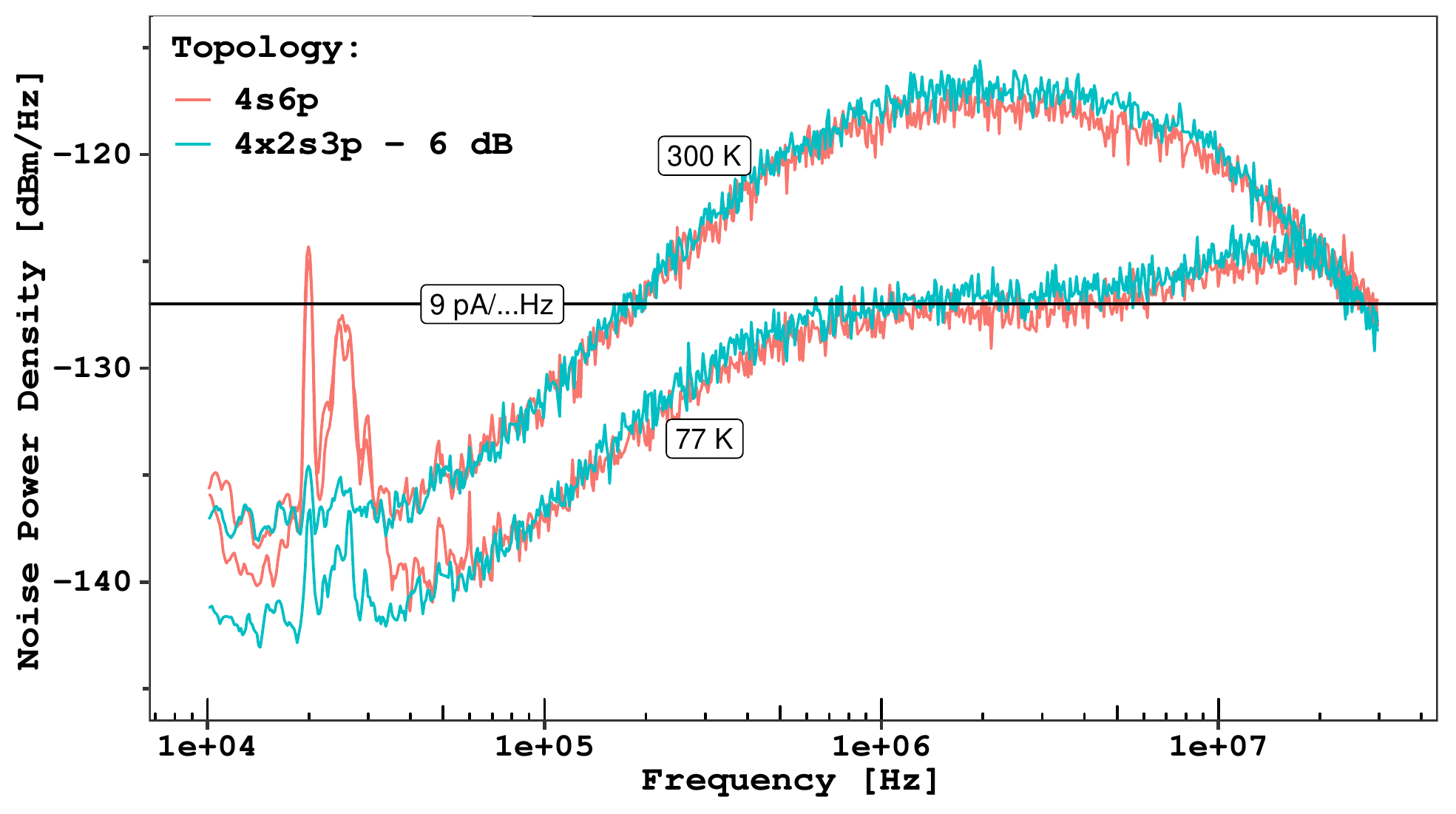}
\caption{Output noise spectra}
\label{fig:noise}
\end{subfigure}
\caption{Trans-impedance amplifier schematics and noise performances.
The output noise density is reported at room temperature and in liquid nitrogen for the same tile configured as 4s6p (red) and as 4$\times$2s3p (blue). The noise of the latter has been scaled by \qty{-6}{\decibel} to simplify the comparison. The horizontal line represents the noise predicted by Equation~\ref{eqn:noise} at \qty{1}{\mega\hertz} for the 4s6p topology. The spurious noise at about \qty{25}{\kilo\hertz} is due to the specific setup that can be configured in both ganging topologies; it is not present in dedicated 4s6p circuits.}
\end{figure}

\subsection{TIA}
We used the trans-impedance amplifier (TIA) described in~\cite{cryo-pre} that is based on the LMH6629 from Texas Instruments operating from \qtyrange{60}{300}{\kelvin}. The chip exhibits a voltage input noise equivalent to a $R_\textrm{TIA}$=\qty{20}{\ohm} resistor and an input current noise ($I_\textrm{TIA}$) that is below \qty{1}{\pico\ampere\per\sqrt{\hertz}} for temperatures lower than \qty{200}{\kelvin}. The TIA is configured with a gain of \qty{10.7}{\kilo\ohm} with disabled compensation and without any feedback capacitor, see Figure~\ref{fig:tia}.

For a generic tile configured with Q quadrants each with P parallel branches each of S SiPMs in series, the total input noise of the TIA in liquid nitrogen at \qty{1}{\mega\hertz} can be calculated with the following formula:
\begin{equation}
I_{n}\{\textrm{1Mhz}\}  =\sqrt{\frac{Q \times  4k_{B} T}{(R_{s}  +  S\times  R_{q}/N_{c}) / P  +  R_\textrm{TIA}} +  2e^{ - }\times Q \times I_\textrm{TIA}}
\label{eqn:noise}
\end{equation}
where the adder noise (if needed) is neglected and $R_s$ is the branch series resistor shown in Figure~\ref{fig:4s6p}.
At \qty{1}{\mega\hertz} the pole $P_\textrm{input}$ at $N_c C_d (R_q / N_c + R_s)$ affects the noise gain by less than \qty{10}{\percent} and therefore the formula for the noise simplifies to a non-inverting amplifier as in Equation~\ref{eqn:noise}.
At lower frequencies, the noise gain is reduced by the presence of the pole $P_\textrm{input}$ and only the $I_\textrm{TIA}$ is relevant. At higher frequencies the parasitic capacitance of the quenching resistor ($C_q \simeq \nicefrac{C_d}{10}$) comes into play leading to an increased noise gain until the cut-off from amplifier bandwidth. Figure~\ref{fig:noise} reports the output noise spectra measured on the same tile configured in 4x2s3p and 4s6p.
At room temperature $R_q$, is smaller by a factor \numrange{2}{3} and the gain bandwidth product of the LMH6629 is one fourth~\cite{cryo-pre}. Therefore, the cut-off happens within the noise gain plateau and spectra does not show any features.

\begin{figure}[tb]
\centering
\begin{subfigure}{.48\textwidth}
\includegraphics[width=\textwidth, ]{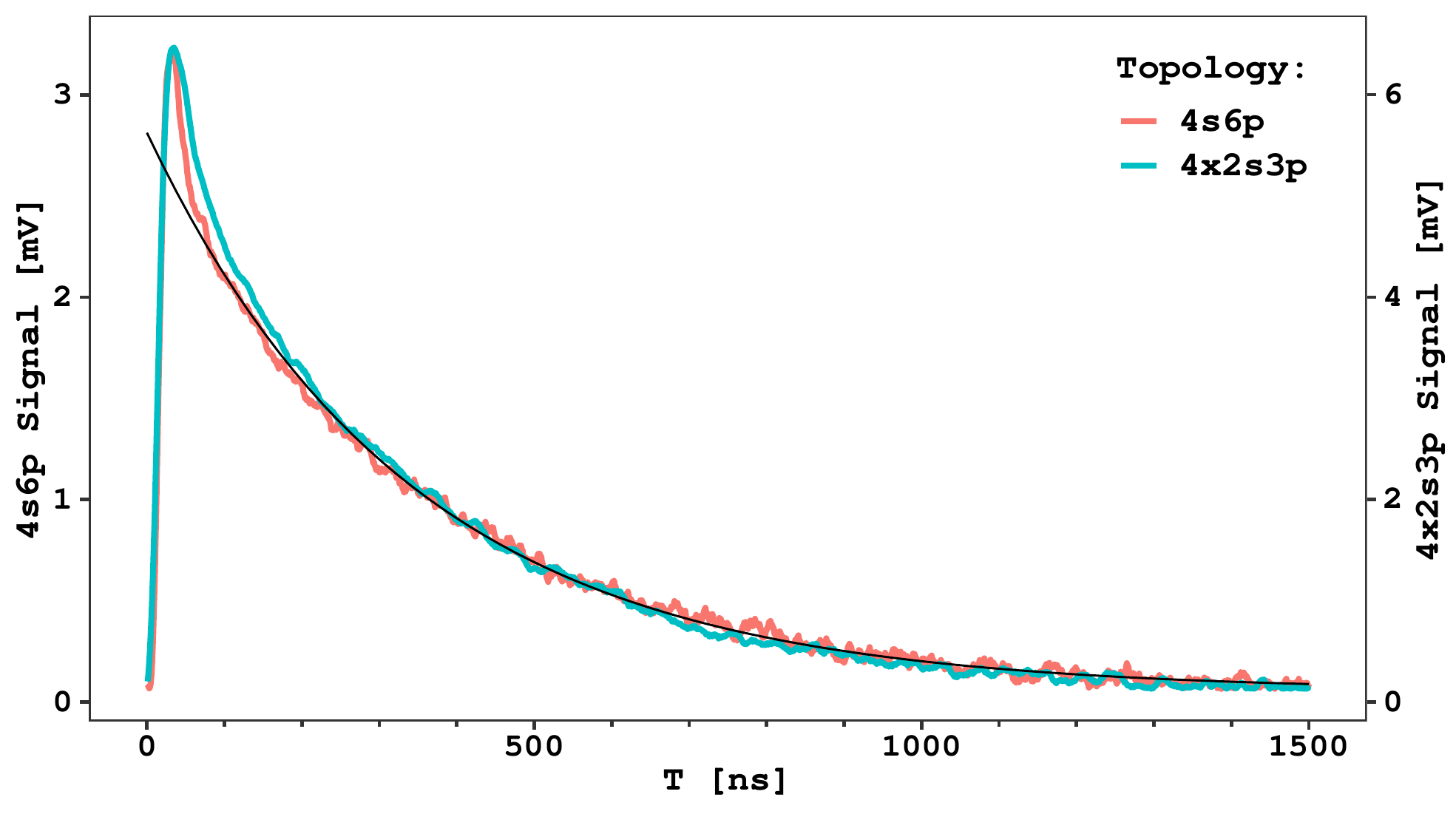}
\caption{Signal shape}
\label{fig:signal}
\end{subfigure}
\hfill
\begin{subfigure}{.48\textwidth}
\includegraphics[width=\textwidth, ]{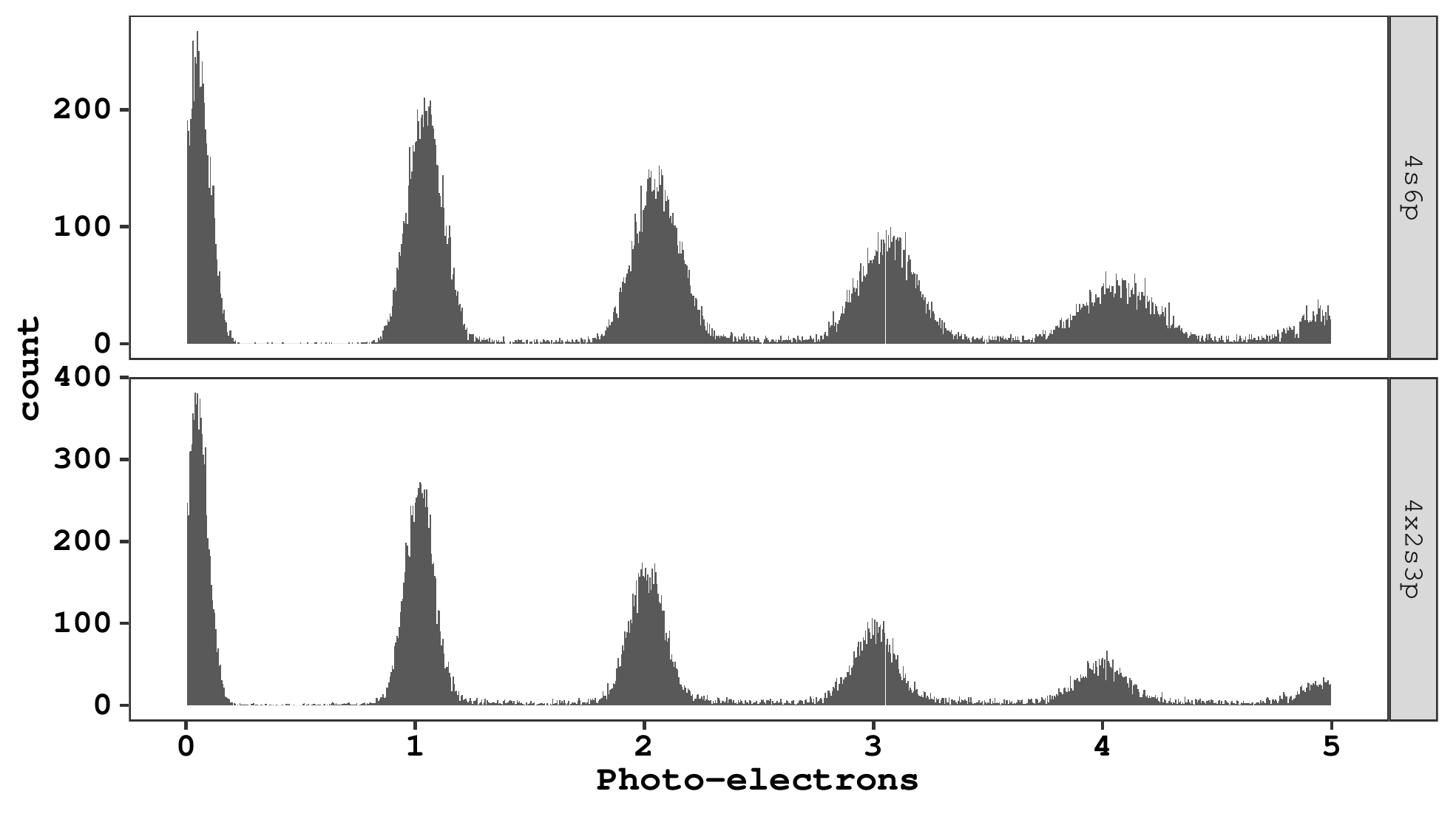}
\caption{Pulse spectrum}
\label{fig:charge}
\end{subfigure}
\caption{Average signal for the first photo-electron of the 4s6p and of the 4x2s3p (scaled by \num{0.5}) topologies in liquid nitrogen at \qty{6.5}{\oV} (left). The 4s6p signal is described as an exponential with $\tau$=\qty{350\pm1}{\nano\second} at better than \qty{50}{\micro\volt} after the first \qty{50}{\nano\second} (black line).
The pulse spectrum for the signal integral (on 3~$\tau$) is reported (right). For both the configurations, SNR larger than \num{30} is achieved over a gain of each SiPM of \num{2.7e6} electrons. The resolution of the first photo-electron peak is \qty{8\pm 1}{\percent} for both topologies.
}
\end{figure}

\begin{figure}[b]
\centering
\includegraphics[width=0.45\textwidth, ]{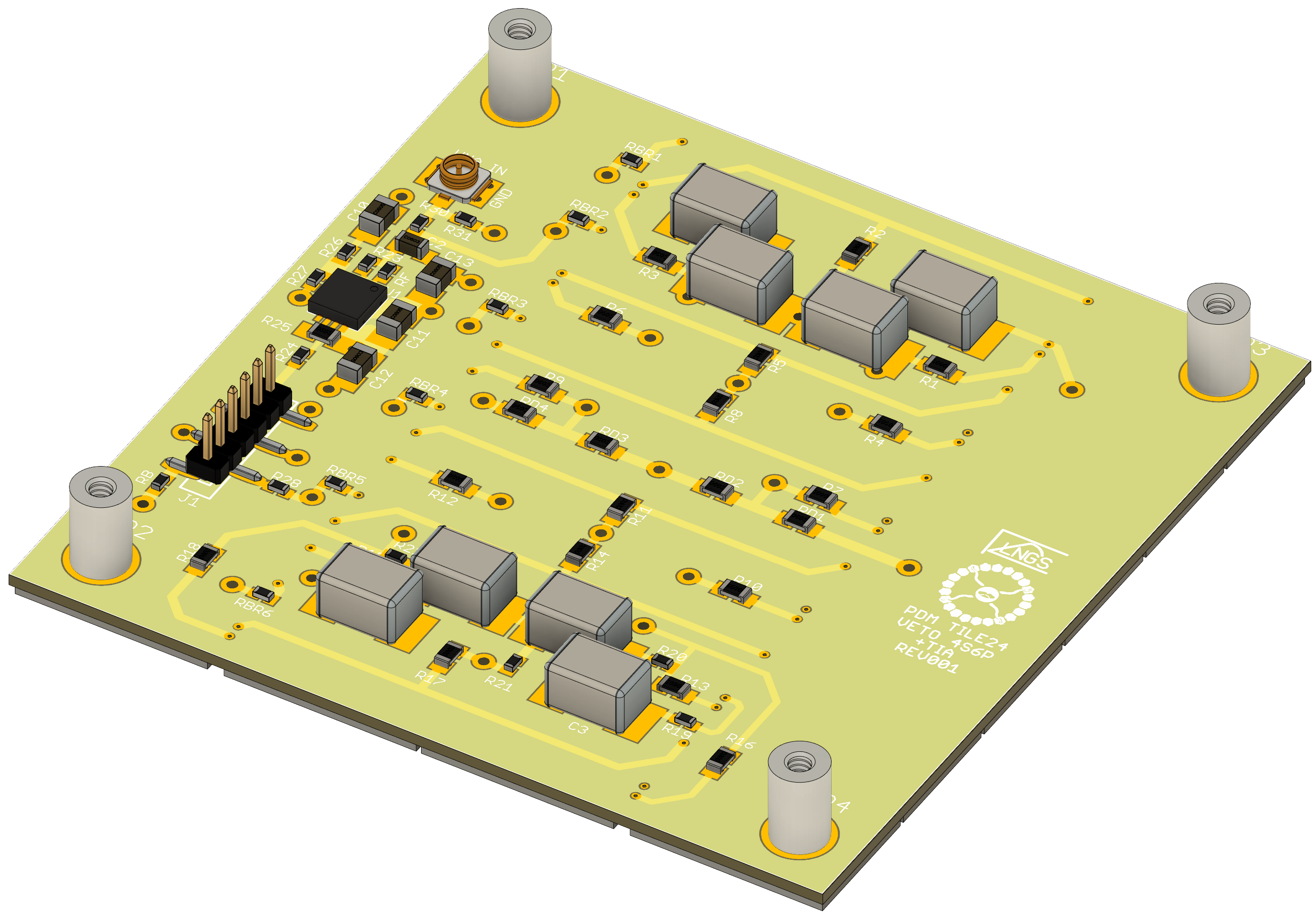}
\hfill
\includegraphics[width=0.45\textwidth, ]{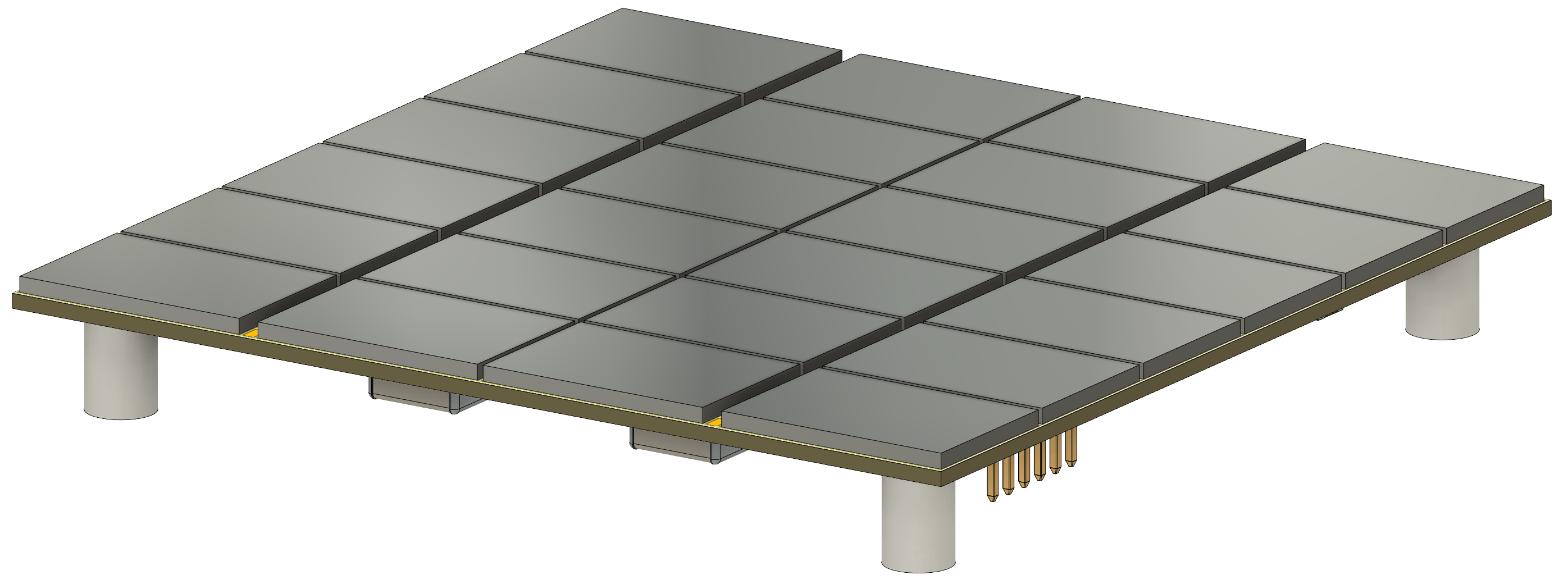}
\caption{Tile+ drawings: the \qty{10}{\pico\farad} capacitor is buried in the PCB. The calibration input, provided by the miniaturized U.FL connector, is useful to test at cryogenic temperature the circuit before bonding the SiPMs.}
\label{fig:tile+:draw}
\end{figure}

\subsection{Pulse Shape and SNR}
As described in~\cite{marano}, the recharge part of the signal of SiPMs includes two exponential components ($\tau_i$ and $\tau_d$). The authors focused on small SiPMs, for which $R_q C_d \gg N R_L (Cd || Cq)$. This leads to the assumption $\tau_i \ll \tau_d$. For \qty{1}{\square\cm} large devices, such assumption breaks and it is necessary to solve the equations with ganged SiPMs. The value of $R_s$=\qty{120}{\ohm} at \qty{77}{\kelvin} leads to the degeneration of the two exponential components, resulting in a pure exponential recharge after the first \qty{50}{\nano\second}, Figure~\ref{fig:signal}.

Concurrently with $R_s$=\qty{120}{\ohm}, the input noise of the 4s6p tile is halved with respect to the noise of the 4x2s3p and $R_s$=\qty{60}{\ohm}, as shown in Figure~\ref{fig:noise}. Similarly, the signal amplitude for the 4s6p is halved, as shown in Figure~\ref{fig:signal}.
Since the signal shape and the noise bandwidth remain unaltered (as result of the identical noise gain), the signal to noise ratio (SNR) is preserved. The SNR is calculated as the gain of the system divided by its average baseline noise (in the pre-trigger window). Figure~\ref{fig:charge} reports the pulse spectrum for the same tile acquired in both configurations. 

\section{Tile+}
A Tile+ is a single unit including 24 SiPMs (top layer) and the required readout electronics (bottom layer). The 4s6p ganging scheme described earlier is used, therefore only one TIA is needed. Figure~\ref{fig:tile+:draw} depict the Tile+. The signal output, the HV bias, the ground and the low voltages (\qty{\pm2.5}{\volt}) are routed on pin strip connector (M50-3630642 from Harwin).

Four threaded \qty{5}{\milli\meter} spacers (97730506330R from Wurth) are soldered to the PCB during the automatic population of bottom components in the first stage of the assembly. 
A low temperature Tin-Bismuth alloy is used to avoid damaging the PEN capacitors which, after reflow at high temperatures, tend to produce leakage in cryogenic environment.

After the validation of the electrical circuit in liquid nitrogen, the SiPMs are placed on the PCB using a Westbond 7200CR manual die bonder.
We use an Indium-Tin solder (Indium Corp Indalloy{\#}1E), dispensed automatically in six \SI{300}{\micro\meter} dots by an Auger valve installed on a fluid dispenser robot~\cite{iza-bari}. As final step, the tile is processed again in the reflow oven with a thermal profile compatible with the bottom components. The PCB finish and the SiPM backside are in gold. This provides good adhesion and avoids the formation of weak inter-metallic compounds with indium.

\begin{figure}[t]
\centering
\includegraphics[width=0.9\textwidth, ]{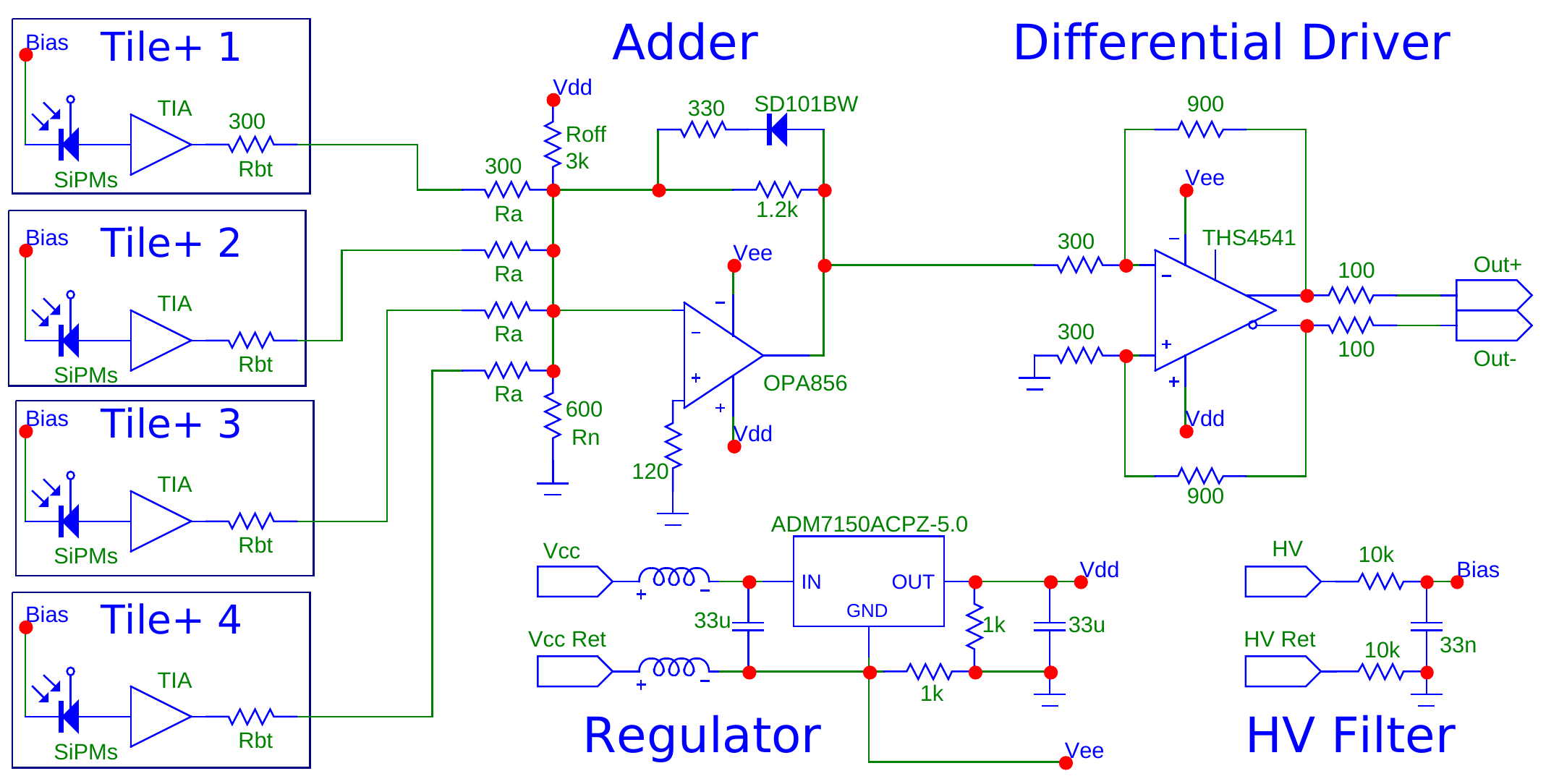}
\caption{Schematics of the MB¼}
\label{fig:mb¼:schem}
\end{figure}

The dimension of the PCB is \qtyproduct{49.5 x 49.5}{\mm} with a fill factor of \qty{90}{\percent}. The PCB is realized in Arlon 55NT that exhibits low thermal shrinkage \qtyrange{6}{9}{ppm\per\kelvin}. This is necessary to avoid stress on the silicon that has a shrinkage of about \qty{2.5}{ppm\per\kelvin}~\cite{iza-bologna}.

In liquid nitrogen the Tile+ dissipates a power of \qty{50}{\milli\watt} with an output swing of \qty{1.7}{\volt} (before back-termination), corresponding to about \num{450} photo-electrons at \qty{7}{\oV}. The performances in terms of SNR, signal shape and pulse spectrum remain unaltered with respect to what is shown in the previous section.

\begin{figure}[t]
\centering
\begin{subfigure}{.48\textwidth}
\includegraphics[width=\textwidth, ]{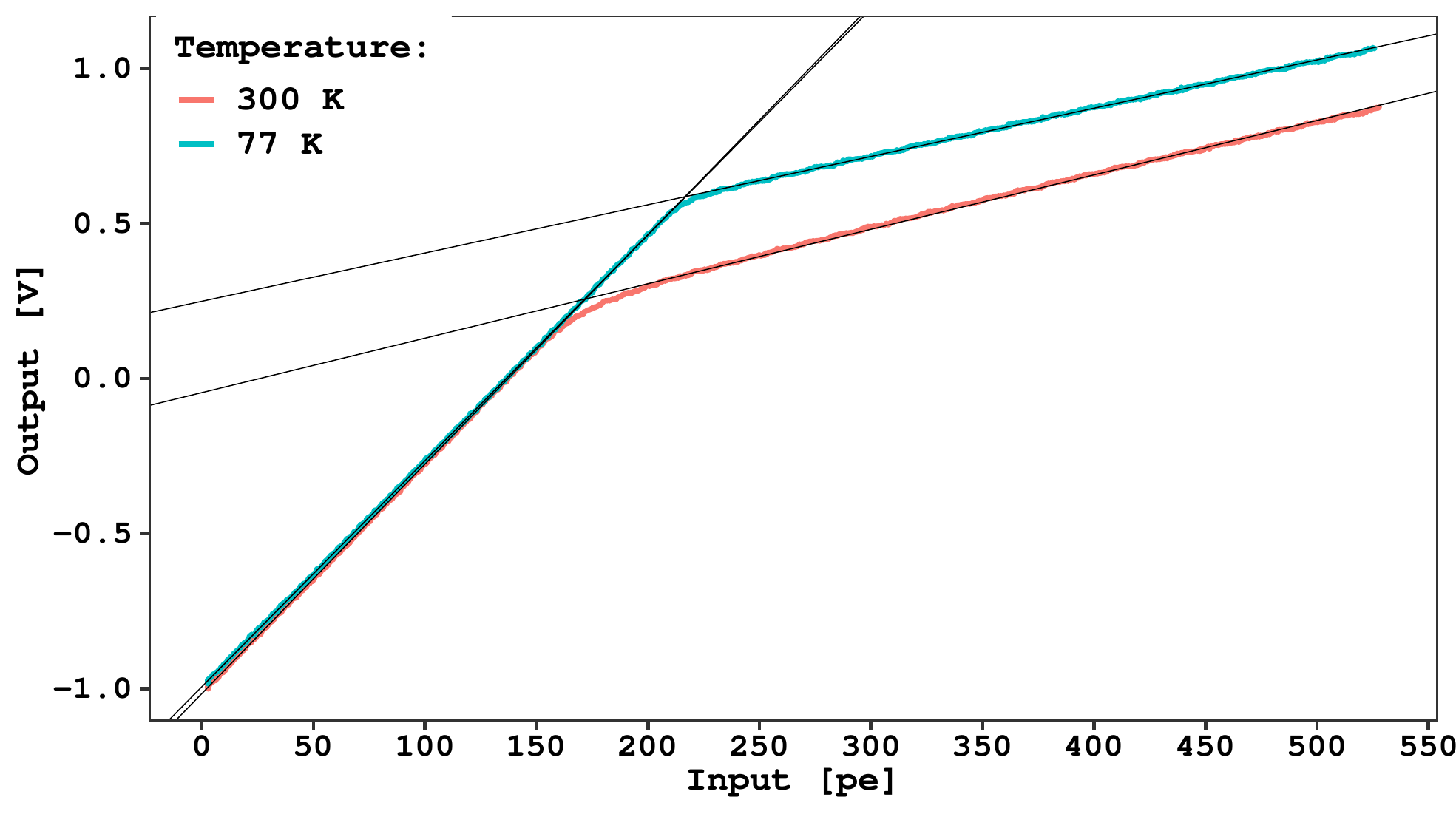}
\caption{Dynamic range}
\label{fig:opa:diode}
\end{subfigure}
\hfill
\begin{subfigure}{.48\textwidth}
\includegraphics[width=\textwidth, ]{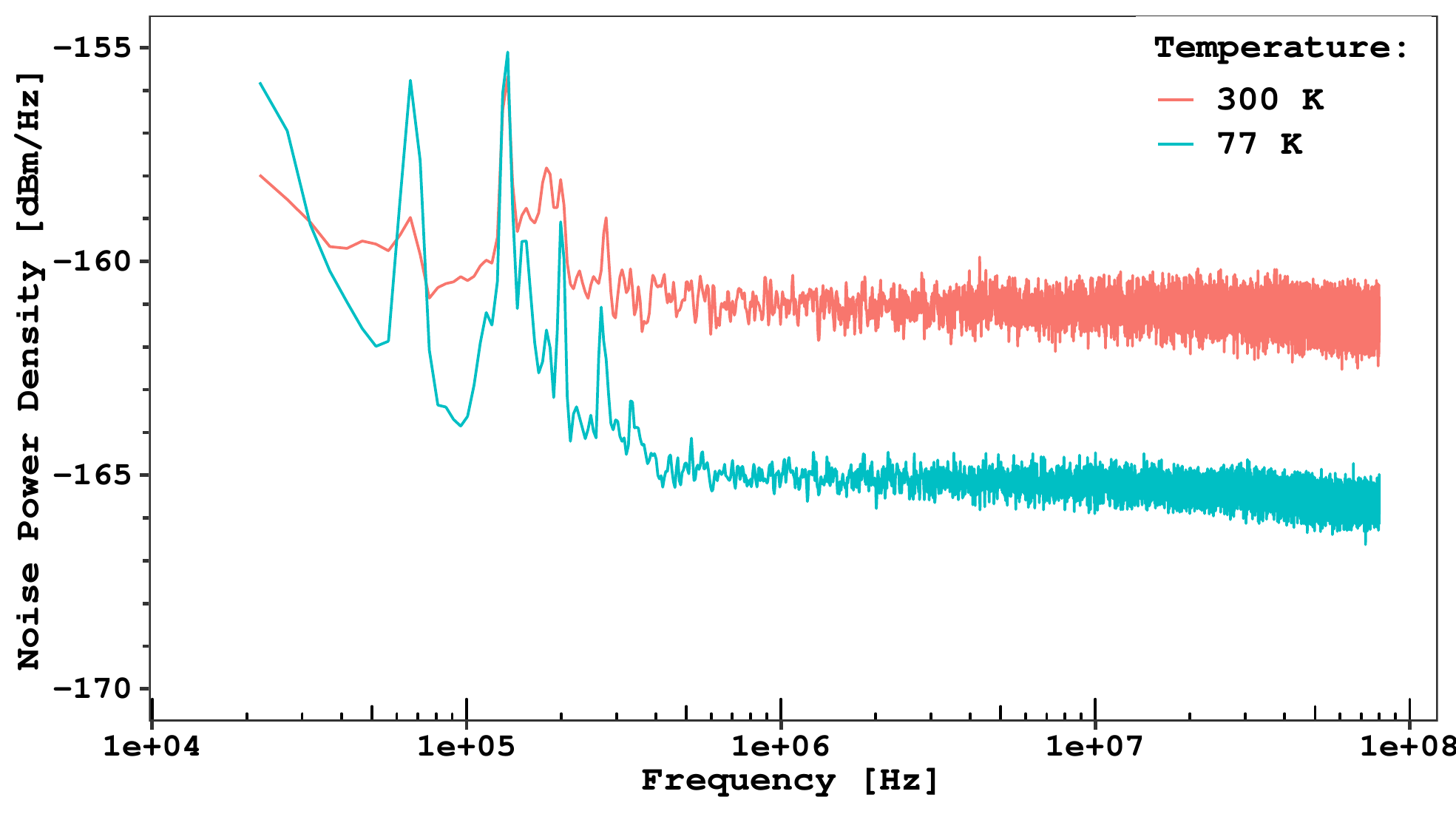}
\caption{Input noise}
\label{fig:opa:noise}
\end{subfigure}
\caption{Perfomances of the OPA856-based adder in liquid nitrogen and at room temperature. For the dynamic range we assume the amplitude of the photo-electron at \qty{3.5}{\milli\volt}. The SD101BW was selected for its low intervention threshold and small distortions at \SI{77}{\kelvin}. Up to the first \num{150} photo-electrons the diode is reverse biased and the system is perfectly linear. 
The input noise spectra of the adder accounts for the noise gain (\num{10}) of the described configuration. Therefore, the right comparison is not with the output noise density of a single TIA, but with the four channels summed (Figure~\ref{fig:noise} scaled by $\sqrt{4}$). The increase of the 1/f noise at cryogenic temperature is typical in the FET-based electronics and becomes relevant only for integration times larger than \qty{.5}{\milli\second}.}
\end{figure}

\section{MB¼}
\begin{figure}[t]
\centering
\begin{minipage}{.45\textwidth}
\includegraphics[width=\textwidth, trim=5mm 15mm 5mm 35mm, clip]{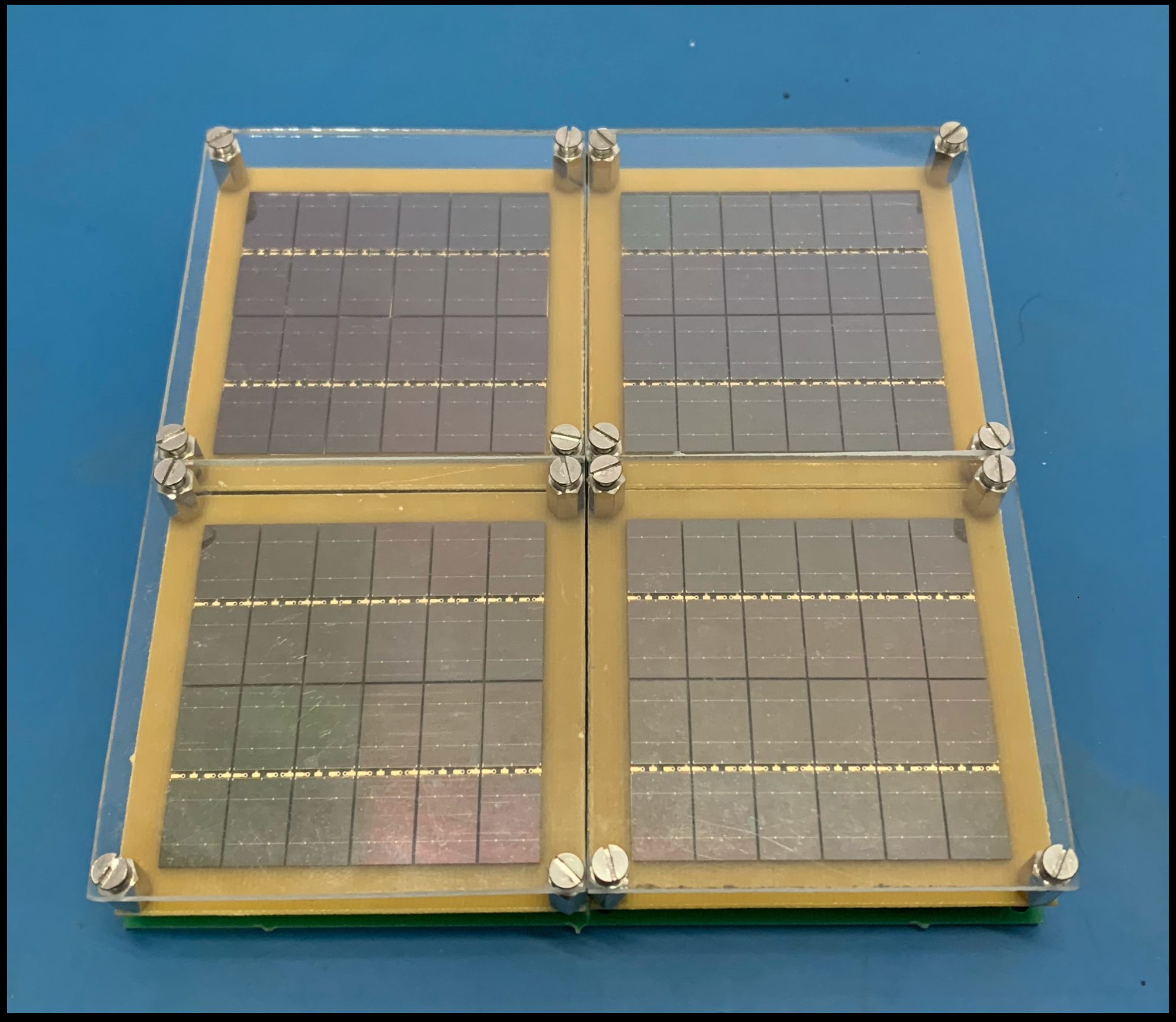}
\end{minipage}
\hskip 1mm
\begin{minipage}{.45\textwidth}
\includegraphics[width=\textwidth, trim=5mm 5mm 5mm 5mm, clip]{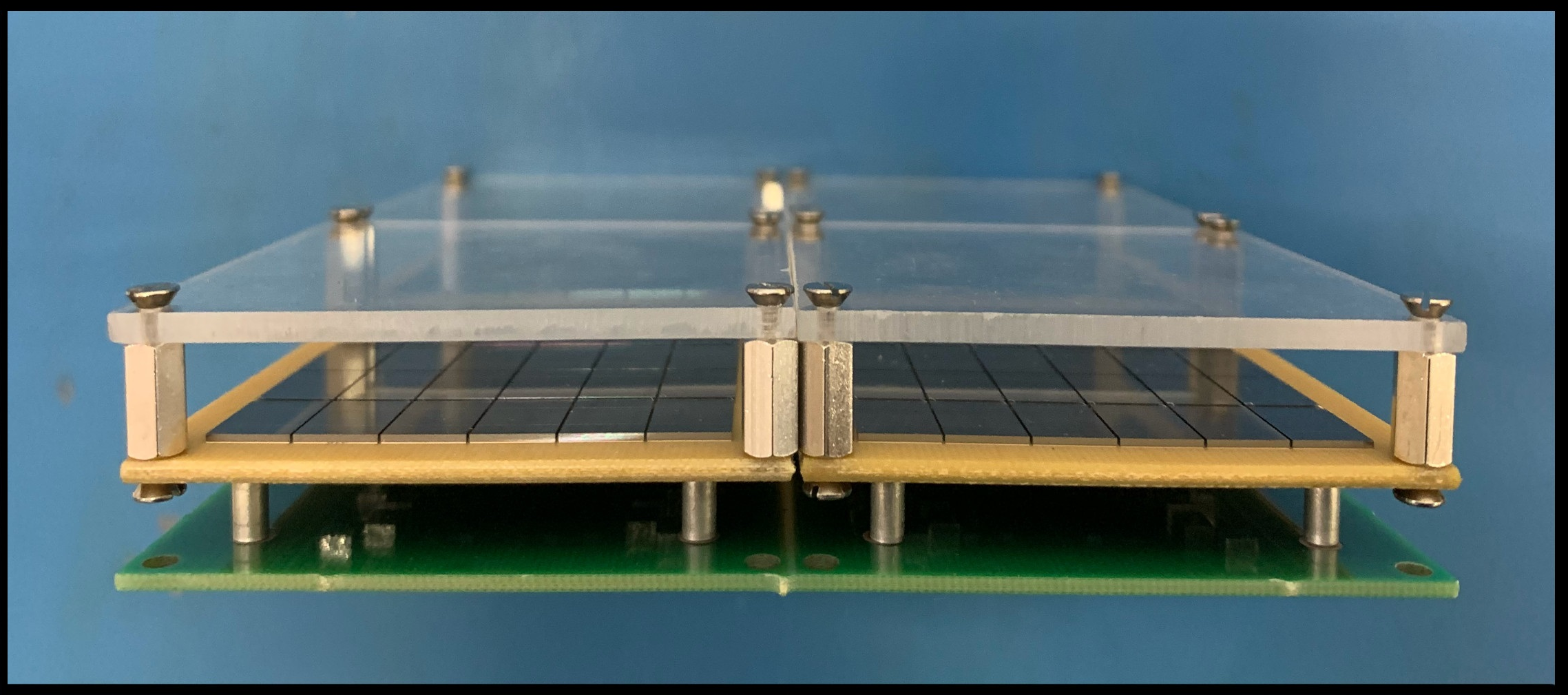}
\vskip 1.5mm
\includegraphics[width=\textwidth, trim=5mm 5mm 5mm 5mm, clip]{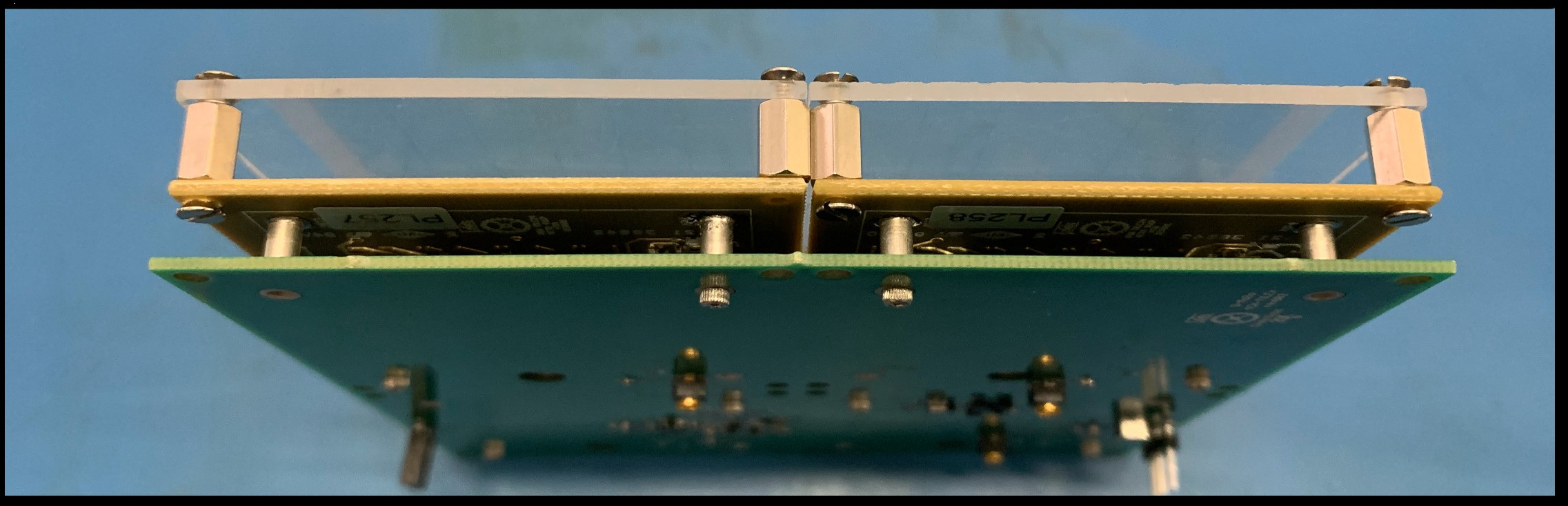}
\end{minipage}
\caption{Populated MB¼: this version has \qty{5}{\mm} frame around each Tile+ to simplify the manipulation and to hold an acrylic protection window. The frame-less version has a fill factor of \qty{88}{\percent} (dimensions \qtyproduct{100x100x7}{\mm}), while this version is at about \qty{60}{\percent} (\qtyproduct{120x120x15}{\mm} with the protection). The PCB of the MB¼ is not critical and either FR4 (as here) or Alron 55NT can be used.}
\label{fig:mb¼:pic}
\end{figure}

\begin{figure}[t]
\centering
\includegraphics[width=0.9\textwidth]{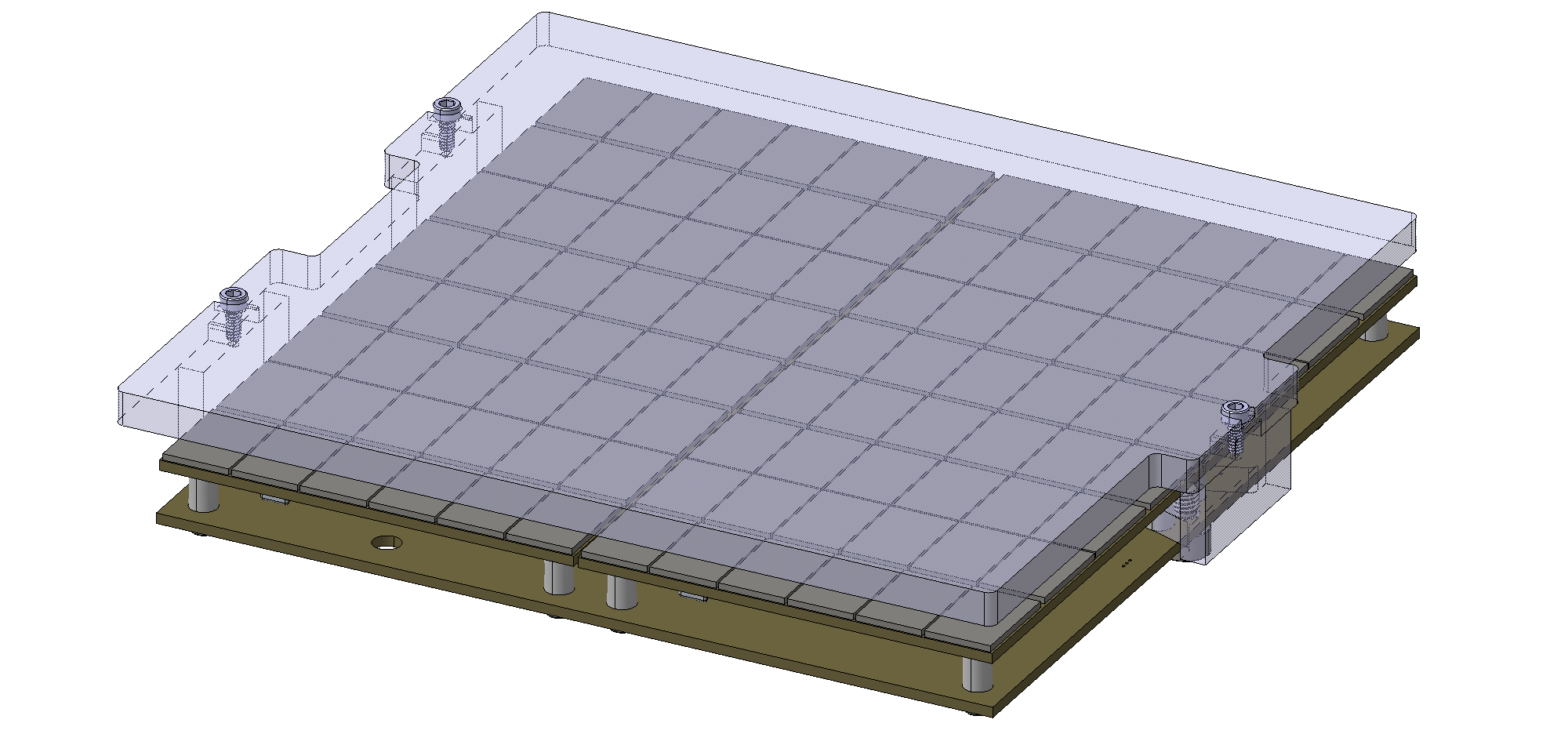}
\caption{Rendering of the MB¼ with an external acrylic protection in a border-less configuration. With this design it is possible to install several MB¼ on an optical plane with only \qty{5}{\mm} of spacing for a total fill factor of \qty{84}{\percent}.}
\label{fig:mb¼:rend}
\end{figure}

Given the very high SNR for the Tile+, it is possible to aggregate four tiles in a single analog photo-detector with a total surface of \qty{100}{\square\cm}. The schematics of such device, called \textit{MB¼}, is shown in Figure~\ref{fig:mb¼:schem}. The finished unit is shown in Figure~\ref{fig:mb¼:pic} and in Figure~\ref{fig:mb¼:rend}.

\subsection{Adder}
The aggregation of the four Tile+ is performed by an analog adder with double gain. This circuit is based on a 
OPA856 from Texas Instruments, capable of working in liquid nitrogen with a gain bandwidth product of about \qty{800}{\mega\hertz}, with a dissipation of \qty{50}{\milli\watt} on a power rail of \qty{5}{\volt}.
At cryogenic temperature the peak to peak output swing is limited to \qty{2.4}{\volt}: to maximise the dynamic range for unipolar (positive) pulses, the output is biased at \qty{-1}{V} by the $R_\textrm{off}$ resistor, leaving a useful swing of \qty{2.2}{\volt}. The small signals gain is \qty{2}{\volt\per\volt}, as defined by $\nicefrac{R_f}{(R_a+R_{bt})}$. The optional Schottky diode (SD101BW) provides a second gain of \qty{.44}{\volt\per\volt} for large pulses. In liquid nitrogen and assuming an over-voltage of \qty{7}{\volt}, without the diode the dynamic range is about 300 photo-electrons, that becomes 500 for the configuration with double gain, see Figure~\ref{fig:opa:diode}.
The bandwidth of the adder at \SI{77}{\kelvin} exceeds \qty{80}{\mega\hertz} and therefore does not affect the pulse shape (the TIA is limited to \qty{30}{\mega\hertz}). Figure~\ref{fig:opa:noise} reports the input noise equivalent for the adder: the comparison term is the output noise of the TIA, Figure~\ref{fig:noise}, incremented by \qty{6}{\decibel}.

\subsection{Differential Driver}
A high dynamic range cryogenic fully differential transmitter has been developed to facilitate scaling up to many photo-detectors in a ground isolated environment (see later). 
The differential transmitter is based on a THS4541 from Texas Instruments. In liquid nitrogen, the chip is capable of driving low impedances (\qtyrange{50}{100}{\ohm} transmission lines) with a large output swing (about \qty{7.4}{\volt} differential) with a supply of \qty{5}{\volt} and a power consumption of \qty{70}{\milli\watt}. To match the output swing of the adder (\qty{2.2}{\volt}), a gain of \qty{3}{\volt\per\volt} per branch is used. Considering the loss of the single-ended to differential conversion (a factor ½), the amplitude of the single photo-electron becomes \qty{10.5}{\milli\volt} before the backside termination (\qty{5}{\milli\volt} on the cable), which is well above the typical noise of a differential line for the \qty{30}{\mega\hertz} bandwidth.
Since the rise-time (\qtyrange[range-phrase=-,range-units=single]{10}{90}{\percent}) of the transmitter with \qty{10}{\meter} of cable is below \qty{4}{\nano\second}
 and its input noise density is in the order of \qty{-160}{dBm}, the effects on the signal quality are negligible.

\subsection{Voltage regulator}
In the Tile+ and in the MB¼, the op-amps are configured as inverting amplifiers. Therefore, all the currents (input and output) sum to zero. As a consequence, it is feasible to create a local ground with a simple voltage divider (and proper bypass capacitors). The main advantage of this design due to the differential transmission, is that it keeps the local ground isolated from the receiver electronics and from the cryostat, which reduces ground loops and noise injection. To further reduce noise and ripple from the power supply, a low drop voltage regulator (LDO) to the system is added. The ADM7150ACPZ-5.0 from Analog Devices provides a protection from the accidental fluctuations of the power supply up to \qty{12}{\volt}, works in liquid nitrogen with noise density in the range of \qtyrange{-160}{-165}{dBm} and a drop of the order of \qty{.5}{\volt}. Therefore, the minimum power consumption of the fully instrumented MB¼ is  \qty{65}{\milli\ampere} x \qty{5.5}{\volt}.

\subsection{Connections}
The MB¼ can be connected with standard unshielded RJ45/U. Categories above 5e satify the bandwidth and the FEP jacket is compatible with cryogenic environment.
At room temperature, two isolated power supplies are required. A Keysight 3649A for the low voltage, and a Keithley 2450 SMU for the HV bias have been used for this purpose.
The HV is internally filtered in the MB¼ with a $\pi$ circuit, that protects the local ground from noise injection.

The signal is delivered to a simple differential receiver implemented with a LMH6552 from Texas Instruments that matches the high dynamic range and low noise requirements. To limit the power dissipation due to the \qty{-1.5}{\volt} of baseline on the termination, an AC coupling on the receiver is required. Standard \qty{100}{\micro\farad} ceramic capacitors with the \qty{100}{\ohm} termination resistor provide a pole well outside the region of interest.

\subsection{Performances}
\begin{figure}[t]
\centering
\begin{subfigure}{.48\textwidth}
\includegraphics[width=\textwidth]{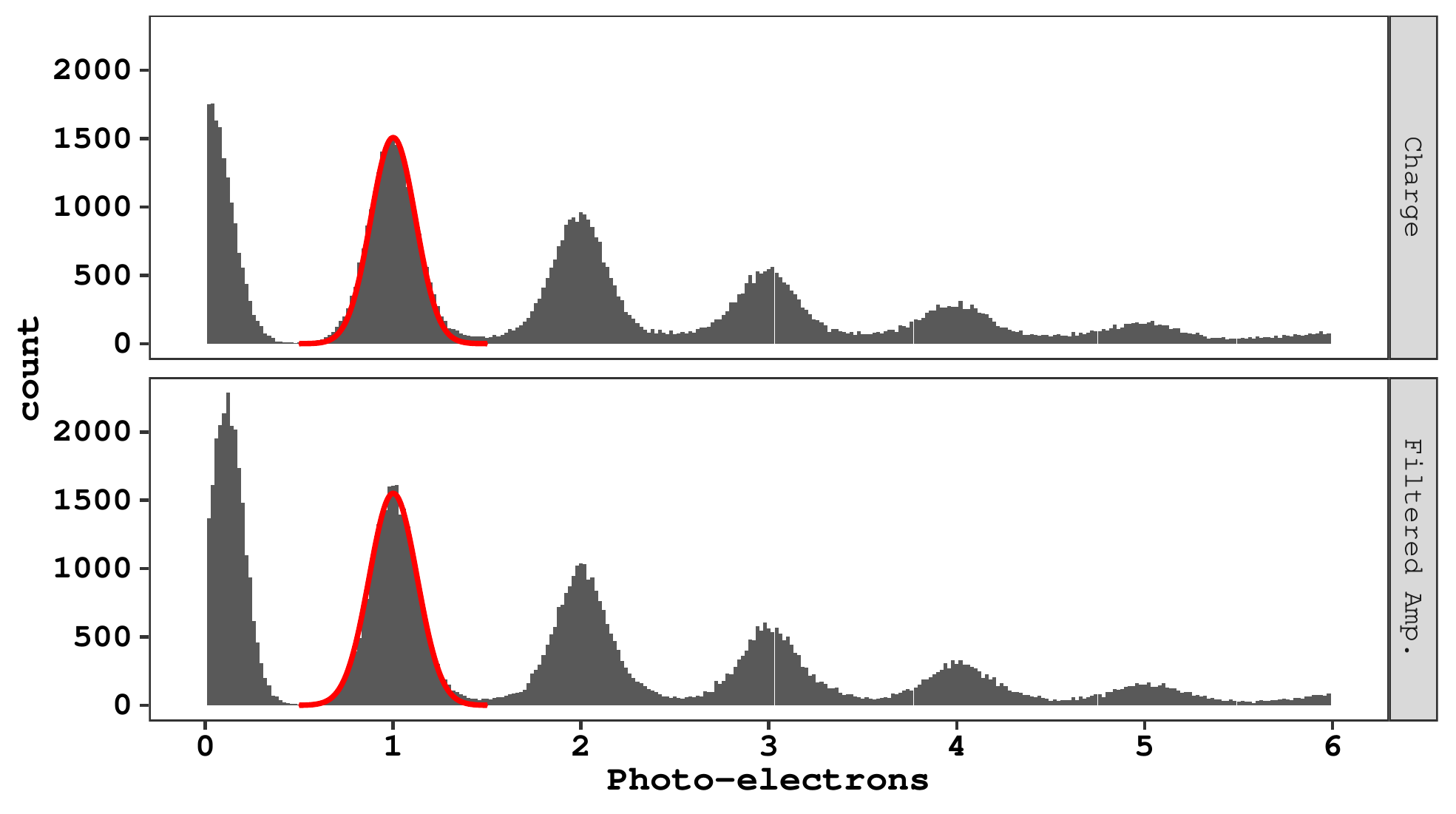}
\caption{Pulse Spectrum}
\label{fig:mb¼:finger}
\end{subfigure}
\hfill
\begin{subfigure}{.48\textwidth}
\includegraphics[width=\textwidth]{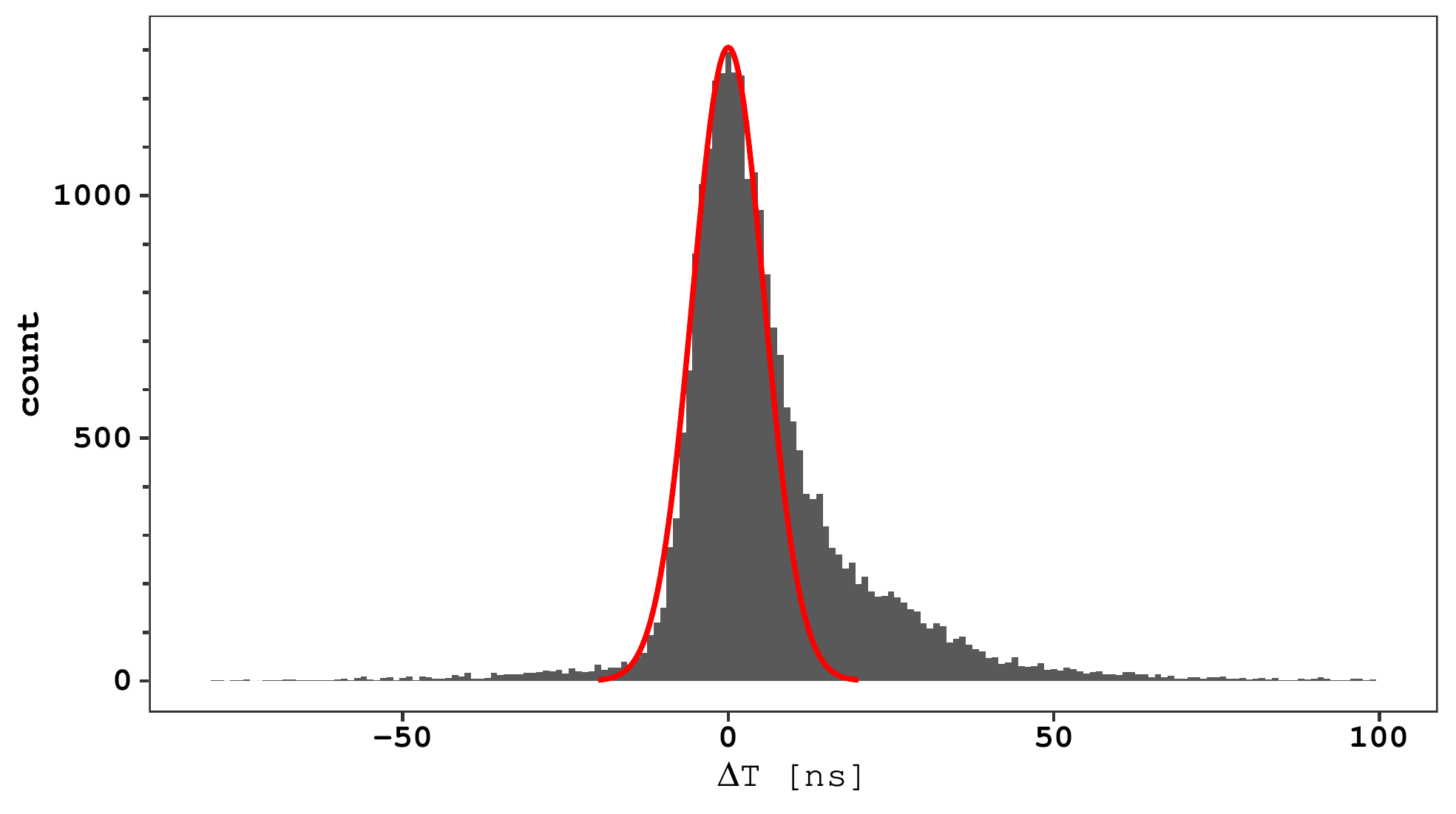}
\caption{Timing}
\label{fig:mb¼:t}
\end{subfigure}
\caption{Performances of the MB¼ at \qty{7}{\oV} in liquid nitrogen. The finger plots for both charge and filtered amplitude (see text for details of filtering) exhibit similar SNR, \num{16\pm 1} versus \num{13.0\pm 0.5}. The resolution of the first photo-electron is \qty{12.5\pm.5}{\percent} for both algorithms.
On the other hand, with the filtered signal it is possible to measure the time of the photo-electrons (relative to the laser pulse) achieving a jitter of \qty{5.5}{\nano\second}. The asymmetric shape of the time jitter is due to the presence of after-pulses in the signal.}
\label{fig:mb¼:res}
\end{figure}

Figure~\ref{fig:mb¼:finger} reports the pulse spectrum of the charge (in a 3~$\tau$ window) and of the filtered amplitude. The latter is based on matched filtering, that is the cross-correlation of raw waveform for the average signal, Figure~\ref{fig:signal}. The result is a cusp-like signal (similar to $e^{-|x|/\tau}$) whose maximum amplitude is used to populate the histogram of Figure~\ref{fig:mb¼:finger}. The location of the maximum provides accurate timing information as shown in Figure~\ref{fig:mb¼:t}. 
The SNR for both charge and filtered signal is above \num{13} and the resolution of the first photo-electron is around \qty{12}{\percent}, dominated by the spread of the breakdown voltage, the disuniformities of the recharge time and the tolerance of the divider.

\section{Conclusions}
In this work, the design and the implementation of a very large SiPM array for cryogenic application has been documented. An SNR larger than \num{13} with a resolution of the first photo-electron \qty{12.5}{\percent} was demonstrated. This is for a photo-detector that includes \num{96}~x~\qty{1}{\square\cm} SiPMs. The timing resolution of this array is \qty{5.5}{\nano\second}, a value better than several PMTs of the same size. The dynamic range exceeds \num{500}~photo-electrons with a power dissipation of \qty{360}{\milli\watt}. The operation of this unit requires standard laboratory power supplies and simple differential receiver.

\section*{Acknowledgements}
We acknowledge support from the Istituto Nazionale di Fisica Nucleare (Italy) and Laboratori Nazionali del Gran Sasso (Italy) of INFN, from NSF (US, Grant PHY-1314507 for Princeton University), from the Royal Society UK and the Science and Technology Facilities Council (STFC), part of the United Kingdom Research.

\bibliographystyle{JHEP}
\bibliography{main, intro}
\end{document}